\documentclass[%
 ,twocolumn%
 ,secnumarabic%
 ,superscriptaddress%
,amssymb, amsmath,nobibnotes, aps, prl,showpacs]{revtex4}
\usepackage{graphicx}
\usepackage{color}
\usepackage{bm}%
\expandafter\ifx\csname package@font\endcsname\relax\else
 \expandafter\expandafter
 \expandafter\usepackage
 \expandafter\expandafter
 \expandafter{\csname package@font\endcsname}%
\fi

\newcommand{\unite}{\mathbf e}
\newcommand{\rc}{r_{\mathrm c}}
\newcommand{\rd}{r_{\mathrm d}}
\newcommand{ \rij}{r_{ij}}
\newcommand{\f}[1]{\mathbf{F}_{#1}}
\newcommand{\fc}[1]{\mathbf{F}^{\mathrm C}_{#1}}
\newcommand{\fr}[1]{\mathbf{F}^{\mathrm R}_{#1}}
\newcommand{\fd}[1]{\mathbf{F}^{\mathrm D}_{#1}}
\newcommand{\unitvec}[1]{\mathbf{e}_{#1}}
\newcommand{\weight}[1]{w_{#1}}
\newcommand{\muint}[1]{A_{#1}}
\newcommand{\pddl}{z}
\newcommand{\slitw}{b}
\newcommand{\timedl}{t}

\newcommand{\visc}{\eta}

\newcommand{\densdpd}{\rho}
\newcommand{\localdens}{\overline{\rho}}

\newcommand{\qleft}{\textquotedblleft}
\newcommand{\qright}{\textquotedblright}
\newcommand{\tc}{\tau_{\mathrm c}}
\newcommand{\stens}[1]{\sigma_{#1}}

\newcommand{\thyd}{\tau_{\mathrm h}}
\newcommand{\tosc}{\tau_{\mathrm osc}}

\newcommand{\capn}{\mathsf {Ca}}

\newcommand{\kb}{k_{\mathrm B}}
\newcommand{\dyncon}{\theta_{\mathrm d}}
\newcommand{\statcon}{\theta_{\mathrm 0}}
\newcommand{\ranampl}{q}

\begin{document}

\title{Nanoscale capillary wetting studied with dissipative particle dynamics}
\author{Claudio Cupelli}
\thanks{Both authors contributed equally to this work.}
\affiliation{
Laboratory for
MEMS applications (Chair Prof. R. Zengerle), Department of Microsystems Engineering(IMTEK),
University of Freiburg, Georges-Koehler-Allee 106, 79110 Freiburg, Germany}
\author{Bj\"orn Henrich}
\thanks{Both authors contributed equally to this work.}
\affiliation{
FMF-Freiburger Materialforschungszentrum, University of Freiburg,
Stefan Meier Str. 21, 79104 Freiburg, Germany}
\author{Michael Moseler}
\affiliation{
FMF-Freiburger Materialforschungszentrum, University of Freiburg,
Stefan Meier Str. 21, 79104 Freiburg, Germany}
\affiliation{
Fraunhofer Institute for Mechanics of Materials, 
W\"ohlerstr. 11, 79108 Freiburg, Germany}
\author{Mark Santer}
\email{santer@imtek.de}
\affiliation{
Laboratory for
MEMS applications (Chair Prof. R. Zengerle), Department of Microsystems Engineering(IMTEK),
University of Freiburg, Georges-Koehler-Allee 106, 79110 Freiburg, Germany%
}
\date{\today}%
\begin{abstract}
We demonstrate that Multi-Body Dissipative Particle Dynamics (MDPD)
can be used as an efficient computational tool for the investigation 
of nanoscale capillary impregnation of confined geometries.  As an 
essential prerequisite, a novel model for a solid-liquid interface 
in the framework of MDPD is introduced, with tunable wetting behaviour 
and thermal roughening to reduce artificial density- and temperature 
oscillations. Within this model, the impregnation dynamics of a 
water-like fluid into a nanoscale slit pore has been studied. Despite 
the coarse graining implied with the model fluid, a sufficient amount 
of non-equilibrium averaging can be achieved allowing for the 
extraction of useful information even from transient simulations, 
such as the dynamic apparent contact angle. Although it is found
to determine the  capillary driving completely, it cannot be intepreted as 
a simple function of the capillary number. 
\end{abstract}

\maketitle
\emph{Introduction.}
Over the last decade, continuous, meso-scale particle simulation methods such as
Dissipative Particle Dynamics (DPD)\cite{hoog92,groot97} and many variants thereof
\cite{espa98a,lowe99,flek00}
have received considerable attention.
Originally invented to include hydrodynamic effects in meso scale simulations of
simple and complex fluids \cite{hoog92,lowe04}, it also has successfully been employed for
studying polymeric systems or melts \cite{groot98,groot99,groot00} or lipid
membranes \cite{groot01}, and for colloidal suspensions \cite{boek96,boek97}.
As a model for solvents, one of the most important recent developments is the introduction of
cohesive properties \cite{pago01,trof02,warr03},
extending the simple quadratic dependance on density in the equation of state (EOS) in early
formulations \cite{groot97}.
The approach of Warren \cite{warr03} leads to particularly
stable liquid-vapor interfaces and could be of great value
in studying free-surface fluid
dynamics problems where thermal capillary fluctuations are important, such as the
intriguing phenomena related to the nanoscale Rayleigh instability
(e.g. the break-up of liquid nano jets \cite{mose00}).
In most numerical studies involving fluid particle (FP)~\cite{FPnote}
methods, the investigations
have largely been kept generic.
Specific interactions, e.g., between solid-liquid interfaces,
have only been accounted for in crude ways,
since in the majority of cases, the
FP-interactions do not arise from a systematic coarse graining procedure starting
at the atomistic scale.
We nevertheless suggest that a FP method can still be suitable for studying
the aforementioned
phenomena, provided that adhesive and interfacial properties are introduced 
carfully and with respect to
the intrinsic molecular characteristics of a given FP model.
It is the purpose of this paper to outline how this could be achieved, 
and we demonstrate for the capillary filling of a narrow slit pore
that detailed information from short term transient simulations 
can be extracted. 

%
%
\emph{General Theory.}
Although the simulation scheme for FP-methods is MD-like in character,
it employs two
features notably different from traditional MD-simulations: soft interaction potentials
of a finite range $\rc$, and a momentum conserving thermostat contributing
a major part to the viscosity of the liquid \cite{lowe99,mast99}.
In DPD, particles interact via pairwise central forces
$\f{ij}=\fr{ij}+\fd{ij}+\fc{ij}$. If $\mathbf {r}_{i}$ denotes the particle position,
the conservative force in standard DPD is
$\fc{ij}=\muint{}\weight{}^{\mathrm C}(r_{ij}) \unitvec{ij}$,
where $\mathbf {r}_{ij}=\mathbf {r}_{i}-\mathbf {r}_{j}$, $r_{ij}=\left| \mathbf {r}_{ij} \right|$
and $\unitvec{ij}=\mathbf {r}_{ij}/r_{ij}$. The weight function $\weight{}^{\mathrm C}(r)=(1-r/r_{c})$
vanishes for an inter-particle distance $r$ larger than a cutoff radius $r_{c}$.
The random and dissipative forces are
$\fr{ij}=\ranampl \weight{}^{\mathrm R}(r_{ij}) \xi_{ij} \mathbf{r}_{ij}$
and $\fd{ij}=-\gamma \weight{}^{\mathrm D}(r_{ij}) ( \mathbf{v}_{ij} \cdot \unitvec{ij}) \unitvec{ij}$,
respectively, and act as a thermostat if the amplitudes $\ranampl$ of the random variable $\xi_{ij}$
and the viscous dissipation $\gamma$ satisfy a fluctuation-dissipation
theorem: $\ranampl^{2}= 2 \gamma \kb T$ and $\weight{ij}^{\mathrm R}(r)^{2}=\weight{ij}^{\mathrm D}(r)$.
The usual choice for the weight functions is $\weight{}^{\mathrm R}=\weight{}^{\mathrm C}$.
Then, the EOS becomes at most quadratic
in the FP-density $\densdpd$, excluding the existence of capillary surfaces. However,
$\fc{ij}$ may be augmented by density dependent contributions \cite{pago01,trof02,warr03}
in order to achieve arbitrary EOS. This class of schemes is termed multi-body DPD (MDPD).
Here, the approach of Warren \cite{warr03} is pursued who has the repulsive part of the force
depend on a weighted average of the particle density, while the attractive part is density independent:
\begin {equation}
\fc{ij}=\muint{ij}\weight{}^{\mathrm C}(\rij)\unite_{ij}
+B(\localdens_i+\localdens_j)\weight{d}(\rij)\unite_{ij},
\end {equation}
with an additional weight function $\weight{}^{\mathrm d}(r)=(1-r/r_{d})^2$.
For a single particle species, $\muint{ij}\equiv \muint{}$ is negative;
the repulsive part with $B>0$ acts at a slightly \emph{smaller} radius of interaction $\rd$.
$\localdens_i$ at the location of particle $i$ is the
instantaneously weighted average $\localdens_i=\sum_{i\neq j}\weight{d}(\rij)$.
Warren showed that this approach produces stable capillary surfaces.
In addition, we introduce a species dependent force
constant $\muint{\mathrm sl}$,
 defining the interaction between liquid(l) particles
and those of the solid(s) wall. The parameters we use for this
model ($A=-40.0$, $B=25.0$) vary slightly from the ones in \cite{warr03}.
The length scale in real space is derived
by matching the compressibility $\partial p/\partial\rho$
of the real fluid (water) to the one
of the model fluid over a particle renormalization factor (the coarse graining
level) $N_m$, as
outlined in \cite{groot01}. Formally, $N_m$ is the number of actual molecules a
single fluid particle is to represent. 
The unit of time is set
equal to $\rc/v_{th}$, where $v_{th}$ means the (real) thermal velocity
of one fluid particle with mass $N_m$ times the molar mass of a
water molecule. This is independent of the transport properties of
the system, as opposed to other prescriptions \cite{groot97}.
All relevant quantities of our MDPD liquid
are summarised in Table \ref{param}.
We stress that the case $N_m=1$ would \emph{not}
correspond to an atomistic limit; the MDPD liquid always remains a
molecular substitute fluid, the molecular characteristics of which 
must be compared with the (continuum) phenomena studied.
\begin{table}[t!]
\caption{\label{param} Parameters used in the simulations.
The surface
tension has been obtained in a planar geometry, by integrating the
lateral part of the pressure tensor in normal direction. The characteristic
fluid parameters are given in model units [m.u.] and in real units (MKS), where
$N_m\simeq3$. Note, that the time step $\Delta t$ for the numerical integration of
the MDPD equations is 2 orders of magnitude larger than $\Delta t$ in a comparable
MD simulation.}
\begin{ruledtabular}
\begin{tabular}{ll|l|ll}
\textbf{Parameter} & \textbf{Symbol} & \textbf{m.u.}& \textbf{MKS}\\\hline
Fluid particle density & $\densdpd$ & $6.00$&$(1000\mathrm{kg}/\mathrm{m}^3)$&\\
Interaction range (attr.) & $\rc$ & $1.0$& $(0.82\mathrm{nm})$&\\
Interaction range (rep.) & $\rd$ & $0.75$&&\\
Amplitude of $\fr{}$& $\ranampl$ & $6.00$&&\\
Compress. &$\partial p/\partial\rho$& $45\pm 2$\\
Surface Tension & $\stens{}$ & $7.51\pm 0.04$ & $(0.138\mathrm{N/m})$&\\
Viscosity & $\visc$ & $7.41\pm 0.06$
\footnote{%
The viscosity was determined by a fit to
a Poiseuille velocity profile for a pressure driven flow in a slit of
width $\slitw$.
}%
&$(0.0003\mathrm{Pa\,s})$&\\
MDPD time step & $\Delta$t & $0.02$ & $76$ fs\\
\end{tabular}
\end{ruledtabular}
\end{table}
\emph{An adequate wall model.}
\label{wall}
\noindent With FP approaches, there are roughly two classes of strategies
to construct solid walls:
(i) use of virtual boundary planes \cite{reve98} and (ii)
walls made of frozen
or crystalline arrangements of fluid particles \cite{clark00,jones99}.
In (i), some rule must be given
to reintroduce a particle to the liquid when it crosses the
boundary, wheras in (ii) the wall acts directly over interparticle
forces.
Models for sharp, rigid interfaces will usually 
produce order effects
and thus density oscillations that on the targeted length
scale are not expected. 
Also, inhomogenous temperature
profiles may arise \cite{fan03}, or other spurious effects (detachment of droplets
under shear flow \cite{jones99}).
In this work, we shall use a thermally roughened wall and
allow fluid particles to penetrate the solid-liquid interface slightly.
To achieve
thermodynamic compatibility,
the solid phase is made of fluid particles at the {\sl same}
density as the fluid, resulting in a strictly homogeneous temperature profile
across the interface. The wall particles are pinned by harmonic forces to their initial
position, and mutually interact via ordinary MDPD forces (eq.~(1) with $\muint{ss}=-40.00$).
An additional, soft harmonic force acting normal
to the interface prevents single fluid particles
from diffusing into the wall indefinetely. 
\begin{figure}[t!]
\includegraphics [width=8.6cm] {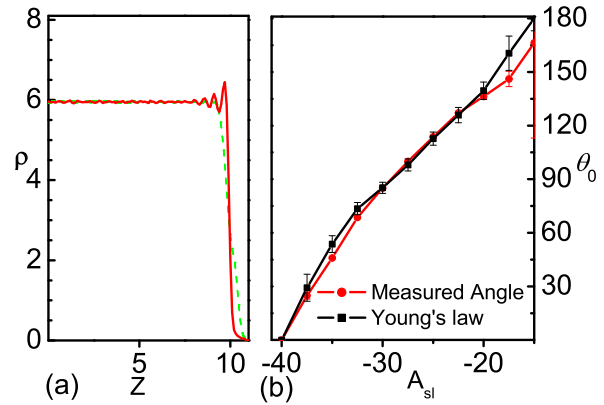}
\caption{\label{fig1:denplot}(color online) (a) Density distribution within a half-slit (full width $20\rc$) for
a mutual attraction strength, $\muint{sl} =-40.0$ for a rigid wall (red line) and thermally roughened wall (green line).
For a rigid wall unphysical density oscillations occur near the wall.
(b) Variation of contact angle with
varying $\muint{sl}$, as determined from a simulated optical measurement ($\bullet$), or using
 Young's law: $\stens{sv} = \stens{sl} + \stens{lv} \cos(\statcon)$,
and obtaining $\stens{sl}$ from integrating
the lateral pressure tensor in planar geometry ($\blacksquare$).
}
\end{figure}
Fig.~\ref{fig1:denplot} illustrates density profiles and the variation of
the static contact angle of this solid-liquid interface.
In contrast to a conventional rigid wall 
density oscillations can be much reduced (compare red and green lines in Fig. 1a, corresponding to 
two different strength of the pinning force). 
Note that the wetting behaviour of the wall can be tuned arbitrarily between hydrophilic
and hydrophobic by diminishing the mutual interaction $\muint{sl}$
between solid- and liquid particles. The interfacial energy can be measured with 
respect to a planar interface, by integrating over the transverse part of the 
pressure tensor; the static contact angle $\statcon$ resulting from the Young-Laplace equation 
is in excellent agreement with an optical measurment (cmp. Fig. 1b), where $\statcon$ is determined
from the meniscus shape of a liquid slab trapped between parallel walls. Details about this model
of a solid-liquid interface will be published elsewhere \cite{henr05}.

\emph{Dynamic behaviour of the impregnation process.}
\label{wick}
In order to demonstrate the applicability of our interface model to nanoscale
impregnation, we study the dynamics of a water-like liquid in
a slit pore of width $b=20$ nm $\simeq 24\rc$  attached to a finite reservoir (see Fig.~2a
for technical  details). 
We consider the case where $\statcon=0$, (i.e. $\muint{sl}=-40.0$).
Fig.~2b and c depict snapshots
of the meniscus at two different times during impregnation.
Partly due to pronounced capillary oscillations that are excited along
with the release in free energy during impregnation, 
the meniscus deviates from a spherical shape.
By ignoring contributions of the wedges, an apparent 
dynamic contact angle
$\dyncon(t)$ can be extracted by a fit to a circle segment (red dashed lines in Fig. 2b,c). 
The evolution of $\cos(\dyncon(t))$ and the penetrated height $\pddl(t)$ over time (red solid lines in 
Fig. 2d, e) can be divided into
two regimes. While for $t<$ 0.35 ns the liquid motion 
is dominated by inertia and high velocities, resulting in a rapid
increase of $\theta_d$,
a slow decrease towards the static contact angle can be observed
for later times.
\begin{figure}
\label{sim:scheme}
\includegraphics[width=8.6cm]{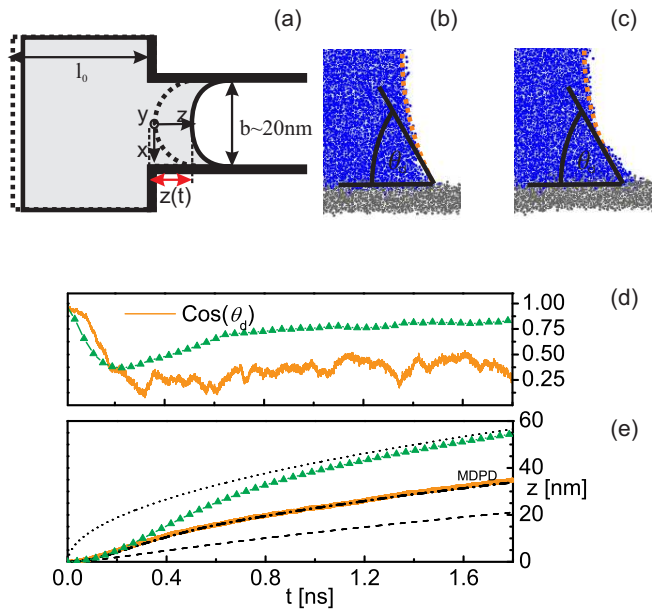}
\caption{\label{wickscheme} (color online) Impregnation of a slit pore. While the simulation cell
is finite in $z-$direction, periodic boundary conditions are applied in $x$ and $y$.
The period in $y$-direction
(slit depth) is set to $12\rc$. The simulation involved $\simeq 170,000$ FP, a complete
run requiring 36 hrs.~ on a single processor workstation. (a) Initially, a spherical
meniscus with the static contact angle $\statcon=0^{\circ}$ has been prepared. Panels b and c show
snapshots of the dynamic contact angle at later times.
Panel d compares the MDPD impregnation dynamics (solid line) with corresppnding
predictions from various continuum models. 
The dash-dotted line results from
an analogous
integration, with the empirical law of Ref.~\cite{brac89}.
Triangles (green) represent $\cos(\dyncon)$
(inset) and $\pddl(\timedl)$, with $\dyncon$ determined from a stationary MDPD plug flow
experiment.
Dashed (green): integration of Eq. \ref{wickdyn}
with driving
Laplace-pressure derived from the
measured dynamic contact angle $\dyncon$, see inset for $\cos(\dyncon(t))$.
As a general reference, the Washburn
solution \cite{wash21}, valid for asymptotically large times ($\dyncon\simeq\statcon$), is
displayed.}
\end{figure}
It is important to note that the spatial dimensions
and time scales probed here (the hydrodynamic time scale $\thyd$ and
the relevant capillary oscillation period $\tosc$) are still well above
the intrinsic molecular scales
of the model fluid \cite{scalenote}. 
This suggests that phenomenologically, 
the impregnation dynamics should follow some continuum law (as it should for a fully
atomistic description) balancing
viscous, capillary and inertial forces
with the momentum change $d/dt(m(t) \dot z(t) + \alpha m_{res}(t) \dot z_{res}(t))$. Here
the time dependent velocities ($\dot z(t)$, $\dot z_{res}(t)$ and accelerated masses ($m(t)$,$m_{res}(t)$)
of the liquid in the slit and a fraction $\alpha$ of the liquid in the reservoir have been introduced.
For a slit geometry as given in Fig.~\ref{wickscheme}, the evolution of the height of the 
miniscus is governed by 
\begin{equation}
\label{wickdyn}
\rho \ddot z(c_{1} \dot z+\alpha l_{0})+ \rho c_{1} {\dot z}^{2} =%
\frac{2\stens {}}{\slitw}\cos(\dyncon)-\frac{12 \eta}{\slitw^2}z \dot z,
\end{equation}
with $c_1=(1-\alpha b/b_{res})$,
where $b/b_{res}$ denotes the ratio
of the slit- to the reservoir width in $x$-direction. 
The Laplace pressure on the right hand of Eq.~2 side is the driving force 
of the impregnation and consequently  
one expects the dynamics solely to be determined by the time dependent
behaviour of the (apparent) dynamic contact angle $\dyncon$. This is indeed the 
case: if the measured course of $\cos(\dyncon(t))$ (red solid line in Fig.~2d) 
is inserted into Eq.~\ref{wickdyn}, a numerical integration results in the 
dashed-dotted line in Fig.~2e, and is in excellent agreement with the numerically
measured $\pddl(t)$, with fluctuations suppressed by viscosity and inertia. 
As we may speak of an average $\dyncon$, it would now be appropriate to make contact 
to standard continuum theories using constitutive laws of the 
form $\dyncon$=$\dyncon(\capn)$, where here $\capn=\eta \dot z/\sigma$ denotes the  
capillary number. The straightforward insertion of empirical laws that have been 
obtained experimentally for a similar range of $\capn$ as found here ($\capn\simeq 0-0.13$)
fails in general (see the law of Joos and Bracke \cite{brac89} as an example, Fig.~2e, dashed line). 
Analysing $\dyncon(\capn)$ in terms of theoretical models such as a chemical or a hydrodynamic
model \cite{broc92} or combinations thereof \cite{gole01}, however, 
is possible but appears not to be very conclusive. 
Rather, it is instructive to
extract a constitutive law $\dyncon(\capn)$ from a  
steady state MDPD simulation with stationary flow profile.
For the relevant range of $\capn$, a sufficiently long fluid plug is 
driven at constant speed $v$
through an infinite slit pore with the same width as in the impregnation
study. The capillary oscillations are much reduced in this setup, and 
the meniscus profile and $\dyncon$ can extracted in a co-moving frame.
If we then insert the resulting $\dyncon(\capn(v=\dot\pddl(t)))$ into 
Eq.~\ref{wickdyn}, the deviation from the measured impregnation is
\emph{again} significant (green triangles in Fig.~2d,e). 

\emph{Discussion and conclusions.}
The apparent \qleft failure\qright of the consistency check above is rooted in the
implicit assumption that in general, the (apparent) dynamic contact angle is a function of 
$\capn$ and possibly a small set of parameters: local 
dissipation mechanisms in a liquid wedge close to or at the triple phase contact 
line are assumed to govern its mobility \cite{broc92}.
However, the realistic numerical setup (afforded by the numerical 
benefits of the FP method)
and the molecular nature 
of our solid-liquid model system include aspects that go beyond a simple picture:
(i) as the simulation is full 3D, the contact line is not only allowed to undulate 
in $x$, but also in $y-$direction. This should affect its average mobility \cite{gole01}. 
(ii)
More important, our results suggests that  $\dyncon$ depends on the
flow field as a whole, as the  Poiseuille profile
of the stationary experiment and the approximate plug flow in the inertial phase of
the impregnation process must be opposed 
\footnote{A difference in flow field  has also been noticed in LJ-simulations 
of sheared liquid slabs,   
with varying $\statcon$; see Gentner, Ogonowski
and De Coninck, Langmuir \textbf{2003}, 19, pp. 3996.}.
A \emph{functional} dependance of $\dyncon$ on the flow field is 
discussed in \cite{blake99};  a key feature is
the dynamics of interface formation, allowing for nonequilibrium values and
gradients in surface and interfacial tension.

It must be emphasized that since the model fluid does not employ atomistic
interactions (e.g., of the LJ-type), the continuum aspects  
of the \qleft true\qright impregnation dynamics are possibly amplified.
Nevertheless, we would like to stress the usefulness of the FP approach as an
efficient tool for the exploration of relevant processes close to and at the
contact line. In combination with subsequent, careful large scale atomistic 
verfication, FP methods can provide important contributions to a unifying picture
of the rich physics occuring in dynamic wetting phenomena.

We gratefully acknowledge the support of Deutsche Forschungsgemeinschaft (DFG),
SPP $1164$, SA 1052/1-1 and Mo 879/4-1.
%
%
%
%
%
%
%
%


\begin{thebibliography}{32}
\bibitem[{FPn()}]{FPnote}
\bibinfo{note}{This terminology is chosen for general continuum, meso-particle
  methods employing unspecific, soft interactions. The shorthand DPD shall
  indicate the special kind of thermostat associated with a given scheme.}
\bibitem[{sca()}]{scalenote}
\bibinfo{note}{The smallest length scale $r_0$ is related to the length scale
  where the density correlation function varies significantly, for the MDPD
  fluid we have $r_0\sim 0.6\rc$. The shortest time scale is determined by the
  decay of the velocity autocorrelation function, representative for a time
  $\tc$ at which single fluid particle collisions take place, we find
  $\tc=0.049\pm0.003$[m.u.]. Continuum behaviour in the sense of generalized
  hydrodynamics can be assumed if, for a characteristic large wave vector $k$
  and a high frequency $\omega$, the inequalities $kr_0\ll 1$ and $\omega\tc\ll
  1$ hold \cite{boon80}. As a rough estimate, one may choose $k=2\pi/\slitw$
  and $\omega=2\pi/\thyd$. $\thyd$ is the hydrodynamic time scale
  $\thyd=\densdpd\slitw^2/12\visc$ on which one expects the transition from the
  inertial to the viscous regime \cite{quer97}. Although the spatial condition
  $kr_0\simeq0.16\ll 1$ seems fair, the value of $\omega\tc\simeq 0.01 \ll 1$
  is satisfactory. As long as there is no excessive local shear gradient, one
  should expect the validity of generalized hydrodynamics, as also conjectured
  for other, atomistic contact angle studies \cite{thom89}. $\tosc$ can be
  estimated from the dispersion relation $\omega^2=(\stens{}/\densdpd)k^3$ to
  be $\simeq 11$[m.u.] (with $k=2\pi/(b/2)$ in $x$-direction)).}
\end{thebibliography}
%
\end{document}